\documentclass[11pt,leqno]{article}

\usepackage{amsmath}
\usepackage{amsthm}
\usepackage{amssymb}
\usepackage[dvips]{graphicx,color}

\theoremstyle{plain}
\newtheorem{theorem}{Theorem}[section]
\newtheorem{lemma}[theorem]{Lemma}
\newtheorem{proposition}[theorem]{Proposition}

\theoremstyle{definition}
\newtheorem{definition}[theorem]{Definition}

\theoremstyle{remark}
\newtheorem{remark}[theorem]{Remark}

\numberwithin{equation}{section}

\begin{document}

\title{\textbf{An operator-theoretical proof for \\the second-order phase transition in \\the BCS-Bogoliubov model of superconductivity}}

\author{Shuji Watanabe\\
Division of Mathematical Sciences\\
Graduate School of Engineering, Gunma University\\
4-2 Aramaki-machi, Maebashi 371-8510, Japan\\
Email: shuwatanabe@gunma-u.ac.jp}

\date{}

\maketitle

\begin{abstract}
%%%%%%%%%%%%%%%%%%%%%%%%%%%%%%%%%%%
We show that the transition from a normal conducting state to a superconducting state is a second-order phase transition in the BCS-Bogoliubov model of superconductivity from the viewpoint of operator theory. Here we have no magnetic field. Moreover we obtain the exact and explicit expression for the gap in the specific heat at constant volume at the transition temperature. To this end, we have to differentiate the thermodynamic potential with respect to the temperature two times. Since there is the solution to the BCS-Bogoliubov gap equation in the form of the thermodynamic potential, we have to differentiate the solution with respect to the temperature two times. Therefore, we need to show that the solution to the BCS-Bogoliubov gap equation is differentiable with respect to the temperature two times as well as its existence and uniqueness. We carry out its proof on the basis of fixed point theorems.
%%%%%%%%%%%%%%%%%%%%%%%%%%%%%%%%%%%

\medskip

Mathematics Subject Classification 2010.  45G10, 47H10, 47N50, 82D55.

Keywords and phrases. BCS-Bogoliubov gap equation, nonlinear integral equation, second-order phase transition, superconductivity.

Running head. An operator-theoretical proof
\end{abstract}

%%%%%%%%%%%%%%%%%%%%%%%%%%%%%%%%%%%%%%%%%%%%%%%%%%%%%%%%%%%%%%%%% 1
\section{Introduction and preliminaries}

In this paper we show that the transition from a normal conducting state to a superconducting state is a second-order phase transition in the BCS-Bogoliubov model of superconductivity from the viewpoint of operator theory. Here we have no magnetic field. Moreover we obtain the exact and explicit expression for the gap in the specific heat at constant volume at the transition temperature. To this end, we have to differentiate the thermodynamic potential (see \eqref{eqn:thermopotential}) with respect to the absolute temperature $T$ two times. Since there is the solution to the BCS-Bogoliubov gap equation in the form of the thermodynamic potential, we have to differentiate the solution with respect to the temperature $T$ two times. Therefore, we need to show that the solution to the BCS-Bogoliubov gap equation is differentiable with respect to the temperature $T$ two times as well as its existence and uniqueness. We carry out its proof on the basis of fixed point theorems.

The BCS-Bogoliubov gap equation \cite{bcs, bogoliubov} is a nonlinear integral equation: 
\begin{equation}\label{eqn:bcsgapeq}
u(T,\,x)=\int_{\varepsilon}^{\hslash\omega_D}
\frac{U(x,\,\xi)\, u(T,\, \xi)}{\,\sqrt{\,\xi^2+u(T,\, \xi)^2\,}\,}\,
\tanh \frac{\,\sqrt{\,\xi^2+u(T,\, \xi)^2\,}\,}{2T}\, d\xi, \quad T \geq 0, \quad \varepsilon \leq x \leq \hslash\omega_D \,,
\end{equation}
where the solution $u$ is a function of the absolute temperature $T$ and the energy $x$. The constant $\omega_D>0$ stands for the Debye angular frequency. The potential $U$ satisfies $U(x,\,\xi)>0$ at all $(x,\,\xi) \in [\varepsilon, \, \hslash\omega_D]^2$.

In \eqref{eqn:bcsgapeq} we need to introduce a cutoff $\varepsilon>0$, which is sufficiently small and fixed. In the original  BCS-Bogoliubov gap equation, one  sets $\varepsilon=0$. However we introduce a very small $\varepsilon>0$.  See Remark \ref{rmk:varepsilon} for the reason why we need to introduce $\varepsilon>0$.

In \eqref{eqn:bcsgapeq} we consider the solution $u$ as a function of the absolute temperature $T$ and the energy $x$. Accordingly, we deal with the integral with respect to the energy $\xi$ in \eqref{eqn:bcsgapeq}. Sometimes one considers the solution $u$ as a function of the absolute temperature and the wave vector. Accordingly, instead of the integral in \eqref{eqn:bcsgapeq}, one deals with the integral with respect to the wave vector over the three dimensional Euclidean space $\mathbb{R}^3$. Odeh \cite{odeh}, and Billard and Fano \cite{billardfano} established the existence and uniqueness of the solution to the BCS-Bogoliubov gap equation for $T=0$, and Vansevenant \cite{vansevenant} for $T \geq 0$. Bach, Lieb and Solovej \cite{bls} studied the gap equation in the Hubbard model for a constant potential, and showed that its solution is strictly decreasing  with respect to the temperature. Frank, Hainzl, Naboko and Seiringer \cite{fhns} studied the asymptotic behavior of the transition temperature (the critical temperature) at weak coupling. Hainzl, Hamza, Seiringer and Solovej \cite{hhss} proved that the existence of a positive solution to the BCS-Bogoliubov gap equation is equivalent to the existence of a negative eigenvalue of a certain linear operator, and showed the existence of a transition temperature. Hainzl and Seiringer \cite{haizlseiringer} obtained upper and lower bounds on the transition temperature and the energy gap for the BCS-Bogoliubov gap equation. For interdisciplinary reviews of the BCS-Bogoliubov model of superconductivity, see Kuzemsky \cite{kuzemsky, kuzemsky2}. See also Kuzemsky \cite[Chapters 26 and 29]{kuzemsky3}.

We define a nonlinear integral operator $A$ by
\begin{equation}\label{eqn:ouroperator}
Au(T,\,x)=\int_{\varepsilon}^{\hslash\omega_D}
\frac{U(x,\,\xi)\, u(T,\, \xi)}{\,\sqrt{\,\xi^2+u(T,\, \xi)^2\,}\,}\,
\tanh \frac{\,\sqrt{\,\xi^2+u(T,\, \xi)^2\,}\,}{2T}\, d\xi.
\end{equation}
Here the right side of this equality is exactly the right side of the BCS-Bogoliubov gap equation \eqref{eqn:bcsgapeq}. Since the solution to the BCS-Bogoliubov gap equation is a fixed point of our operator $A$, we apply fixed point theorems to our operator $A$.

Let $U_1>0$ is a  positive constant and set $\displaystyle{ U(x,\,\xi)=U_1 }$ at all $(x,\,\xi) \in [\varepsilon, \, \hslash\omega_D]^2$. Then the solution to the BCS-Bogoliubov gap equation becomes a function of the temperature $T$ only, and we denote the solution by $\Delta_1$. Accordingly, the BCS-Bogoliubov gap equation \eqref{eqn:bcsgapeq} is reduced to the simple gap equation \cite{bcs}
\begin{equation}\label{eqn:delta1}
1=U_1\int_{\varepsilon}^{\hslash\omega_D}
 \frac{1}{\,\sqrt{\,\xi^2+\Delta_1(T)^2\,}\,}\,
 \tanh \frac{\, \sqrt{\,\xi^2+\Delta_1(T)^2\,}\,}{2T}\,d\xi, \quad 0 \leq T \leq\tau_1 \, ,\end{equation}
where the temperature $\tau_1>0$ is defined by (see \cite{bcs})
\[
1=U_1\int_{\varepsilon}^{\hslash\omega_D}
\frac{1}{\,\xi\,}\,\tanh \frac{\xi}{\,2\tau_1\,}\,d\xi.
\]
See also Niwa \cite{niwa} and Ziman \cite{ziman}.

As is well known, physicists and engineers studying superconductivity always assume that there is a unique nonnegative solution $\Delta_1$ to the simple gap equation \eqref{eqn:delta1}, that the solution $\Delta_1$ is continuous and strictly decreasing with respect to the temperature $T$, and that the solution $\Delta_1$ is of class $C^2$ with respect to the temperature $T$, and so on. But, as far as the present author knows, there is no mathematical proof for these assumptions imposed in the BCS-Bogoliubov model. Applying the implicit function theorem to the simple gap equation \eqref{eqn:delta1}, we obtain the following proposition that indeed gives a mathematical proof for these assumptions:
\begin{proposition}[{\cite[Proposition 1.2]{watanabe1}}]\label{prp:delta-one}
Let $U_1>0$ is a  positive constant and set $U(x,\,\xi)=U_1$ at all $(x,\,\xi) \in [\varepsilon, \, \hslash\omega_D]^2$. Set
\[
\Delta=\frac{\, \sqrt{\, \left(\hslash\omega_D-\varepsilon \, e^{1/U_1} \right)\left(\hslash\omega_D -\varepsilon \, e^{-1/U_1}\right) \, } \,}{\,\sinh\frac{1}{\,U_1\,}\,}.
\]
Then there is a unique nonnegative solution $\Delta_1: [\,0,\,\tau_1\,] \to [0,\,\infty)$ to the simple gap equation \eqref{eqn:delta1} such that the solution $\Delta_1$ is continuous and strictly decreasing with respect to the temperature $T$ on the closed interval $[\,0,\,\tau_1\,]$:
\[
\Delta_1(0)=\Delta>\Delta_1(T_1)>\Delta_1(T_2)>\Delta_1(\tau_1)=0, \qquad 0<T_1<T_2<\tau_1.
\]
Moreover, the solution $\Delta_1$ is of class $C^2$ with respect to the temperature $T$ on the interval $[\,0,\,\tau_1\,)$ and satisfies
\[
\Delta_1'(0)=\Delta_1''(0)=0 \quad \mbox{and} \quad \lim_{T\uparrow \tau_1} \Delta_1'(T)=-\infty.
\]
\end{proposition}

\begin{remark}
We set $\Delta_1(T)=0$ at $T>\tau_1$. See figure 1.
\end{remark}

We then introduce another positive constant $U_2>0$. Let $0<U_1<U_2$ and set $U(x,\,\xi)=U_2$ at all $(x,\,\xi) \in [\varepsilon,\, \hslash\omega_D]^2$. Then a similar discussion implies that for $U_2$, there is a unique nonnegative solution $\Delta_2: [\,0,\,\tau_2\,] \to [0,\,\infty)$ to the simple gap equation
\begin{equation}\label{eqn:delta2}
1=U_2\int_{\varepsilon}^{\hslash\omega_D}
 \frac{1}{\,\sqrt{\,\xi^2+\Delta_2(T)^2\,}\,}\,
 \tanh \frac{\, \sqrt{\,\xi^2+\Delta_2(T)^2\,}\,}{2T}\,d\xi, \qquad
0\leq T\leq \tau_2\, .
\end{equation}
Here, $\tau_2>0$ is defined by
\[
%\begin{equation}\label{eqn:tau2}
1=U_2\int_{\varepsilon}^{\hslash\omega_D}
\frac{1}{\,\xi\,}\,\tanh \frac{\xi}{\,2\tau_2\,}\,d\xi.
%\end{equation}
\]

\begin{remark}
We again set $\Delta_2(T)=0$ at $T>\tau_2$.
\end{remark}

\begin{lemma}[{\cite[Lemma 1.5]{watanabe1}}] \quad {\rm (a)} The inequality $\tau_1<\tau_2$ holds. \\
\noindent {\rm (b)} If \   $0\leq T<\tau_2$, then $\Delta_1(T)<\Delta_2(T)$. If \  $T\geq \tau_2$, then $\Delta_1(T)=\Delta_2(T)=0$.
\end{lemma}
See figure 1.  The function $\Delta_2$ has properties similar to those of the function $\Delta_1$.

%%%%%%%%%%%%%%%%%%%%%%%%%%%
\begin{figure}[htbp]
\includegraphics[width=16cm]{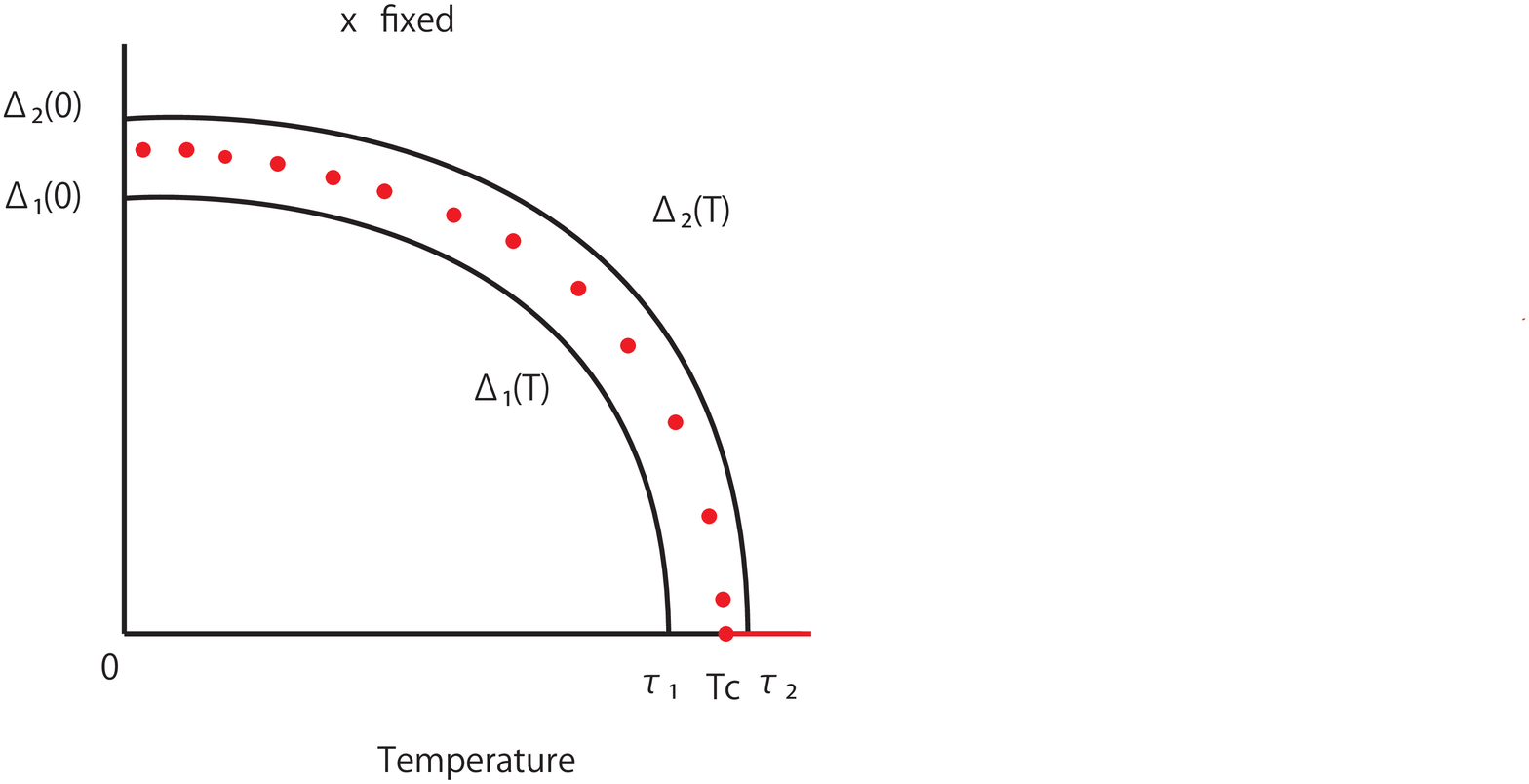}
\caption{\textsf{The graphs of the functions $\Delta_1$ and $\Delta_2$ with $x$ fixed. The solution $u_0(T,\, x)$ is between $\Delta_1(T)$ and $\Delta_2(T)$ for each $T$.}}
\end{figure}
%%%%%%%%%%%%%%%%%%%%%%%%%%%

Let us turn to the BCS-Bogoliubov gap equation \eqref{eqn:bcsgapeq}. We assume the following condition on $U$:
\begin{equation}\label{eqn:conditionU}
U(\cdot,\,\cdot) \in C([\varepsilon,\, \hslash\omega_D]^2), \qquad (0<) \; U_1 \leq U(x,\,\xi) \leq U_2 \quad \mbox{at all} \quad (x,\,\xi) \in [\varepsilon,\, \hslash\omega_D]^2.
\end{equation}
Let $0 \leq T \leq \tau_2$ and fix $T$. We now consider the Banach space $C[0,\, \hslash\omega_D]$ consisting of continuous functions of the energy $x$ only, and deal with the following temperature dependent subset $V_T$:
\[
V_T=\left\{ u(T,\,\cdot) \in C[\varepsilon,\, \hslash\omega_D]: \; \Delta_1(T) \leq u(T,\,x) \leq \Delta_2(T) \; \mbox{at} \; x \in [\varepsilon,\, \hslash\omega_D] \right\}.
\]

\begin{remark}
The set $V_T$ depends on the temperature $T$. See figure 1.
\end{remark}

The following theorem gives another proof of the existence and uniqueness of the nonnegative solution to the BCS-Bogoliubov gap equation, and shows how the solution varies with the temperature.
\begin{theorem}[{\cite[Theorem 2.2]{watanabe1}}] \label{thm:3-1}
Assume \eqref{eqn:conditionU} and let $T \in [0,\, \tau_2]$ be fixed. Then there is a unique nonnegative solution $u_0(T,\,\cdot) \in V_T$ to the BCS-Bogoliubov gap equation \eqref{eqn:bcsgapeq}:
\[
u_0(T,\, x)=\int_{\varepsilon}^{\hslash\omega_D}
\frac{U(x,\,\xi)\, u_0(T,\, \xi)}{\,\sqrt{\,\xi^2+u_0(T,\, \xi)^2\,}\,}\,
\tanh \frac{\,\sqrt{\,\xi^2+u_0(T,\, \xi)^2\,}\,}{2T}\, d\xi, \, \quad  x \in [\varepsilon,\, \hslash\omega_D].
\]
Consequently, the solution $u_0(T,\,\cdot)$ with $T$ fixed is continuous with respect to the energy $x$ and varies with the temperature as follows:
\[
\Delta_1(T) \leq u_0(T,\, x) \leq \Delta_2(T) \quad \mbox{at} \quad
(T,\,x) \in [0,\, \tau_2] \times [\varepsilon,\, \hslash\omega_D].
\]
\end{theorem}

See figure 1 for the graph of the solution $u_0$ with the energy $x$ fixed.

\begin{remark}
Let $u_0(T,\,\cdot)$ be as in Theorem \ref{thm:3-1}. If there is a point $x_1 \in [\varepsilon,\, \hslash\omega_D]$ satisfying $u_0(T,\, x_1)=0$, then $u_0(T,\, x)=0$ at all $x \in [\varepsilon,\, \hslash\omega_D]$. See \cite[Proposition 2.4]{watanabe1}.
\end{remark}

The existence and uniqueness of the transition temperature $T_c$ were pointed out in previous papers
\cite{fhns, hhss, haizlseiringer, vansevenant}. In our case, we can define it as follows:
\begin{definition}\label{dfn:tc}
Let $u_0(T,\,\cdot)$ be as in Theorem \ref{thm:3-1}. Then the transition temperature $T_c$ is defined by 
\[
T_c=\inf\{ T>0: \, u_0(T,\, x)=0 \quad \mbox{at all} \quad x \in [\varepsilon,\, \hslash\omega_D] \}.
\]
\end{definition}

\begin{remark}\label{rmk:tau1Tc}
Let $u_0(T,\,\cdot)$ be as in Theorem \ref{thm:3-1}. We then set $u_0(T,\, x)=0$ at all $x \in [\varepsilon,\, \hslash\omega_D]$ and at $T \geq T_c$ . The transition temperature $T_c$ is the critical temperature that divides normal conductivity and superconductivity, and satisfies $\tau_1\leq T_c \leq \tau_2$ . See figure 1.
\end{remark}

But Theorem \ref{thm:3-1} tells us nothing about continuity of the solution $u_0$ with respect to the temperature $T$. Applying the Banach fixed-point theorem, we then showed in \cite[Theorem 1.2]{watanabe2} that the solution $u_0$ is indeed continuous both with respect to the temperature $T$ and with respect to the energy $x$ under the restriction that the temperature $T$ is sufficiently small. See also \cite{watanabe3}.

In order to discuss the second-order phase transition we need to deal with the thermodynamic potential, as mentioned before. Let us introduce the thermodynamic potential $\Omega$ in the BCS-Bogoliubov model without the magnetic field:
\[
\Omega=-T \ln Z,
\]
where $Z$ denotes the partition function. Throughout this paper we use the unit $k_B=1$. Generally speaking, the thermodynamic potential $\Omega$ is a function of the temperature $T$, the chemical potential and the volume of our physical system under consideration. However we fix both the chemical potential and the volume of our physical system, and so we consider the thermodynamic potential $\Omega$ as a function of the temperature $T$ only. We have only to deal with the difference $\Psi$ between the thermodynamic potential corresponding to superconductivity and that corresponding to normal conductivity. The difference $\Psi$ of the thermodynamic potential in the BCS-Bogoliubov model is given by
\begin{eqnarray}\label{eqn:thermopotential}
\Psi(T)
&=& -2N_0 \int_{\varepsilon}^{\hslash\omega_D} \left\{ \sqrt{\,\xi^2+u_0(T,\, \xi)^2\,}-\xi \right\} \, d\xi \\
& & +N_0 \int_{\varepsilon}^{\hslash\omega_D} \frac{u_0(T,\, \xi)^2}{\,\sqrt{\,\xi^2+u_0(T,\, \xi)^2\,}\,}\,
\tanh \frac{\,\sqrt{\,\xi^2+u_0(T,\, \xi)^2\,}\,}{2T}\, d\xi \nonumber \\
& & -4N_0 T \int_{\varepsilon}^{\hslash\omega_D} \ln
\frac{\, 1+e^{ -\sqrt{\,\xi^2+u_0(T,\, \xi)^2\,}/T } \,}{  1+e^{-\xi/T}  } \, d\xi, \quad T \in [\tau, \, T_c],
 \nonumber \\ \nonumber
\end{eqnarray}
where $N_0$ stands for the density of states per unit energy at the Fermi surface, and $u_0$ is the solution to the BCS-Bogoliubov gap equation \eqref{eqn:bcsgapeq}. Here, $\tau$ is that in Theorem \ref{thm:solution}, and $T_c$ is the transition temperature. We define the difference $\Psi$ only on the interval $[\tau, \, T_c]$ because we are interested in the phase transition at the transition temperature $T_c$~.

\begin{definition}\label{dfn:thermodynamic}
The transition from a normal conducting state to a superconducting state at $T=T_c$ is a second-order phase transition if the difference $\Psi$ of the thermodynamic potential satisfies the following:

\smallskip
\noindent (a) \    $\Psi \in C^2[\tau,\, T_c]$ \   and \   $\Psi(T_c)=0$.

\smallskip
\noindent (b) \    $\displaystyle{ \frac{\, \partial \Psi \,}{\partial T}(T_c)=0  }$.

\smallskip
\noindent (c) \    $\displaystyle{ \frac{\, \partial^2 \Psi \,}{\partial T^2}(T_c) \not= 0  }$.
\end{definition}

\begin{remark}\label{rmk:gapcv}
Condition (a) of Definition \ref{dfn:thermodynamic} implies that the thermodynamic potential $\Omega$ is continuous at an arbitrary temperature $T$. Conditions (a) and (b) imply that the entropy $S=-(\partial \Omega/\partial T)$ is also continuous at an arbitrary temperature $T$ and that, as a result, no latent heat is observed at $T=T_c$ . Hence Conditions (a) and (b) imply that the transition at $T=T_c$ is not a first-order phase transition. On the other hand, Conditions (a) and (c) imply that the specific heat at constant volume $C_V=-T\, (\partial^2 \Omega/\partial T^2)$ is discontinuous at $T=T_c$ and that the gap $\Delta C_V$ in $C_V$ is observed at $T=T_c$ . Here, the gap $\Delta C_V$ at $T=T_c$ is given by
\[
\Delta C_V=-T_c \, \frac{\, \partial^2 \Psi \,}{\partial T^2}(T_c).
\]
For more details on the entropy and the specific heat at constant volume, see e.g. \cite[Section III]{bcs} or Niwa \cite[Section 7.7.3]{niwa}.
\end{remark}

\begin{remark}\label{rmk:varepsilon}
When we differentiate the difference $\Psi$ given by \eqref{eqn:thermopotential} with respect to $T$, we have, for example, the term
\begin{equation}\label{eq:term}
-N_0 \, \int_{\varepsilon}^{\hslash\omega_D} \frac{1}{\, \sqrt{\,\xi^2+u_0(T,\, \xi)^2\,} \,} \left\{ \frac{\partial}{\,\partial T \,} \, u_0(T,\, \xi)^2 \right\} \, d\xi.
\end{equation}
Note that $u_0(T_c\,,\, \xi)=0$ at all $\xi$ and that
\[
\left. \frac{\partial}{\,\partial T \,} \, u_0(T,\, \xi)^2 \right|_{T=T_c} = -v(\xi) < 0
\]
at all $\xi$. Here the function $v$ is that in Condition (C) of Section 2. The term \eqref{eq:term} then becomes, at $T=T_c$,
\[
N_0 \, \int_{\varepsilon}^{\hslash\omega_D} \frac{\, v(\xi)\,}{\, \xi \,} \, d\xi.
\]
When $\varepsilon=0$, we find that the term diverges at $T=T_c$ without any assumption on the function $v$. Moreover, if the potential $U$ is a constant, then the solution to the BCS-Bogoliubov gap equation \eqref{eqn:bcsgapeq} depends on the temperature $T$ only, and does not depend on the energy $\xi$ (see Proposition \ref{prp:delta-one}). So the term \eqref{eq:term} becomes
\begin{equation}\label{eqn:termconst}
-N_0 \,  \left\{ \frac{\partial}{\,\partial T \,} \, u_0(T)^2 \right\} \, \int_{\varepsilon}^{\hslash\omega_D} \frac{1}{\, \sqrt{\,\xi^2+u_0(T)^2\,} \,} \, d\xi.
\end{equation}
Note that $u_0(T_c)=0$ and that
\[
\left. \frac{\partial}{\,\partial T \,} \, u_0(T)^2 \right|_{T=T_c} = -v.
\]
Here, $v$ is a constant, and it is assumed frequently that $v>0$ in the BCS-Bogoliubov model. The term \eqref{eqn:termconst} then becomes, at $T=T_c$,
\[
N_0 \, v \, \int_{\varepsilon}^{\hslash\omega_D} \frac{1}{\, \xi \,} \, d\xi.
\]
When $\varepsilon=0$, we again find that the term diverges at $T=T_c$. This is why we need to introduce $\varepsilon>0$ both in the BCS-Bogoliubov gap equation \eqref{eqn:bcsgapeq} and in the difference $\Psi$ given by \eqref{eqn:thermopotential}.
\end{remark}

%%%%%%%%%%%%%%%%%%%%%%%%%%%%%%%%%%%%%%%%%%%%%%%%%%%%%%%%%%%%%%% 2
\section{Main results}

Let the potential $U(\cdot,\,\cdot)$ satisfy the following:
\begin{equation}\label{eqn:conditionUprime}
U(\cdot,\,\cdot) \in C([\varepsilon,\, \hslash\omega_D]^2), \qquad (0<) \; U_1 < U(x,\,\xi) < U_2 \quad \mbox{at all} \quad (x,\,\xi) \in [\varepsilon,\, \hslash\omega_D]^2.
\end{equation}
Then, by Theorem \ref{thm:3-1}, there is a unique nonnegative solution $u_0(T,\,\cdot) \in V_T$ to the BCS-Bogoliubov gap equation \eqref{eqn:bcsgapeq}. By Definition \ref{dfn:tc}, the transition temperature $T_c>0$ is thus defined. Note that the transition temperature $T_c>0$ is related to the  solution $u_0(T,\,\cdot) \in V_T$.

The function
\[
(T,\, x) \mapsto \int_{\varepsilon}^{\hslash\omega_D} \frac{U(x,\,\xi)}{\,\sqrt{\,\xi^2+\Delta_2(T)^2\,}\,}\,
 \tanh \frac{\, \sqrt{\,\xi^2+\Delta_2(T)^2\,}\,}{2T}\,d\xi
\]
is continuous and its value is less than $1$. This is because
\begin{eqnarray*}
& & \int_{\varepsilon}^{\hslash\omega_D} \frac{U(x,\,\xi)}{\,\sqrt{\,\xi^2+\Delta_2(T)^2\,}\,}\,
 \tanh \frac{\, \sqrt{\,\xi^2+\Delta_2(T)^2\,}\,}{2T}\,d\xi \nonumber \\
&<& \int_{\varepsilon}^{\hslash\omega_D} \frac{U_2}{\,\sqrt{\,\xi^2+\Delta_2(T)^2\,}\,}\,
 \tanh \frac{\, \sqrt{\,\xi^2+\Delta_2(T)^2\,}\,}{2T}\,d\xi \nonumber \\
&=&1
\end{eqnarray*}
by \eqref{eqn:delta2}. For example, if the potential $U(x.\,\xi)$ is nearly equal to $0.8 \, U_2$, then
\[
\int_{\varepsilon}^{\hslash\omega_D} \frac{U(x,\,\xi)}{\,\sqrt{\,\xi^2+\Delta_2(T)^2\,}\,}\,
 \tanh \frac{\, \sqrt{\,\xi^2+\Delta_2(T)^2\,}\,}{2T}\,d\xi
\]
is nearly equal to $0.8$. Note that the function
\[
(T,\, x) \mapsto \int_{\varepsilon}^{\hslash\omega_D}
  \frac{\, U(x,\,\xi) \,}{\xi}\, \tanh \frac{\xi}{\, 2T \,}\,  d\xi 
\]
is also continuous.

We choose suitable $\tau>0$ and $\varepsilon>0$ such that $\tau<T_c$ and
\begin{eqnarray}\label{eqn:the-inequality}
& &\int_{\varepsilon}^{\hslash\omega_D} \frac{U(x,\,\xi)}{\,\sqrt{\,\xi^2+\Delta_2(T)^2\,}\,}\,
 \tanh \frac{\, \sqrt{\,\xi^2+\Delta_2(T)^2\,}\,}{2T}\,  d\xi \\ \nonumber
& & \quad +\frac{\, \Delta_2(\tau)^2 \,}{2\, \varepsilon^2}  \int_{\varepsilon}^{\hslash\omega_D}
  \frac{\, U(x,\,\xi) \,}{\xi}\, \tanh \frac{\xi}{\, 2T \,}\,  d\xi <1. \nonumber
\end{eqnarray}
The first term on the left side of \eqref{eqn:the-inequality} is less that 1 as mentioned above. The second term tends to $0$ as $\Delta_2(\tau)/ \varepsilon \to 0$ since
\begin{eqnarray*}
 \, & & \frac{\, \Delta_2(\tau)^2 \,}{2\, \varepsilon^2}  \int_{\varepsilon}^{\hslash\omega_D}
  \frac{\, U(x,\,\xi) \,}{\xi}\, \tanh \frac{\xi}{\, 2T \,}\,  d\xi \\
  &<& \frac{\, \Delta_2(\tau)^2 \,}{2\, \varepsilon^2} \, U_2 \,\ln \frac{\, \hslash\omega_D \,}{\varepsilon} \\
  &=& \frac{\, \Delta_2(\tau)^2 \,}{2\, \varepsilon^2} \, U_2 \,
  \ln \left\{ \cosh \frac{1}{\, U_2\,}+\sqrt{\, 1+\frac{\, \Delta_2(0)^2 \,}{\, \varepsilon^2} \,}
   \;  \sinh \frac{1}{\, U_2\,}  \right\}. \\
 &\to& 0 \quad \mbox{as} \quad \frac{\, \Delta_2(\tau)\,}{\varepsilon} \to 0.
\end{eqnarray*}
Here we used the equality (see Proposition \ref{prp:delta-one})
\[
\Delta_2(0)=\frac{\, \sqrt{\, \left(\hslash\omega_D-\varepsilon \, e^{1/U_2} \right)\left(\hslash\omega_D -\varepsilon \, e^{-1/U_2}\right) \, } \,}{\,\sinh\frac{1}{\,U_2\,}\,}.
\]

\begin{remark}
The function $\Delta_2(T)$ is strictly decreasing with respect to $T$ and tends to $0$ as $T \to \tau_2$ , while $\varepsilon>0$ is fixed and is not equal to $0$. Therefore, there is a certain $\tau>0$ satisfying $\Delta_2(T)<\varepsilon$ for $T \in [\tau,\, T_c]$. See figure 2. Hence $\Delta_2(\tau)/ \varepsilon <1$. Thus we can choose suitable $\tau>0$ and $\varepsilon>0$ such that the inequality \eqref{eqn:the-inequality} holds true.
\end{remark}

%%%%%%%%%%%%%%%%%%%%%%%%%%%
\begin{figure}[htbp]\hspace{-1.5cm}
\includegraphics[width=14cm]{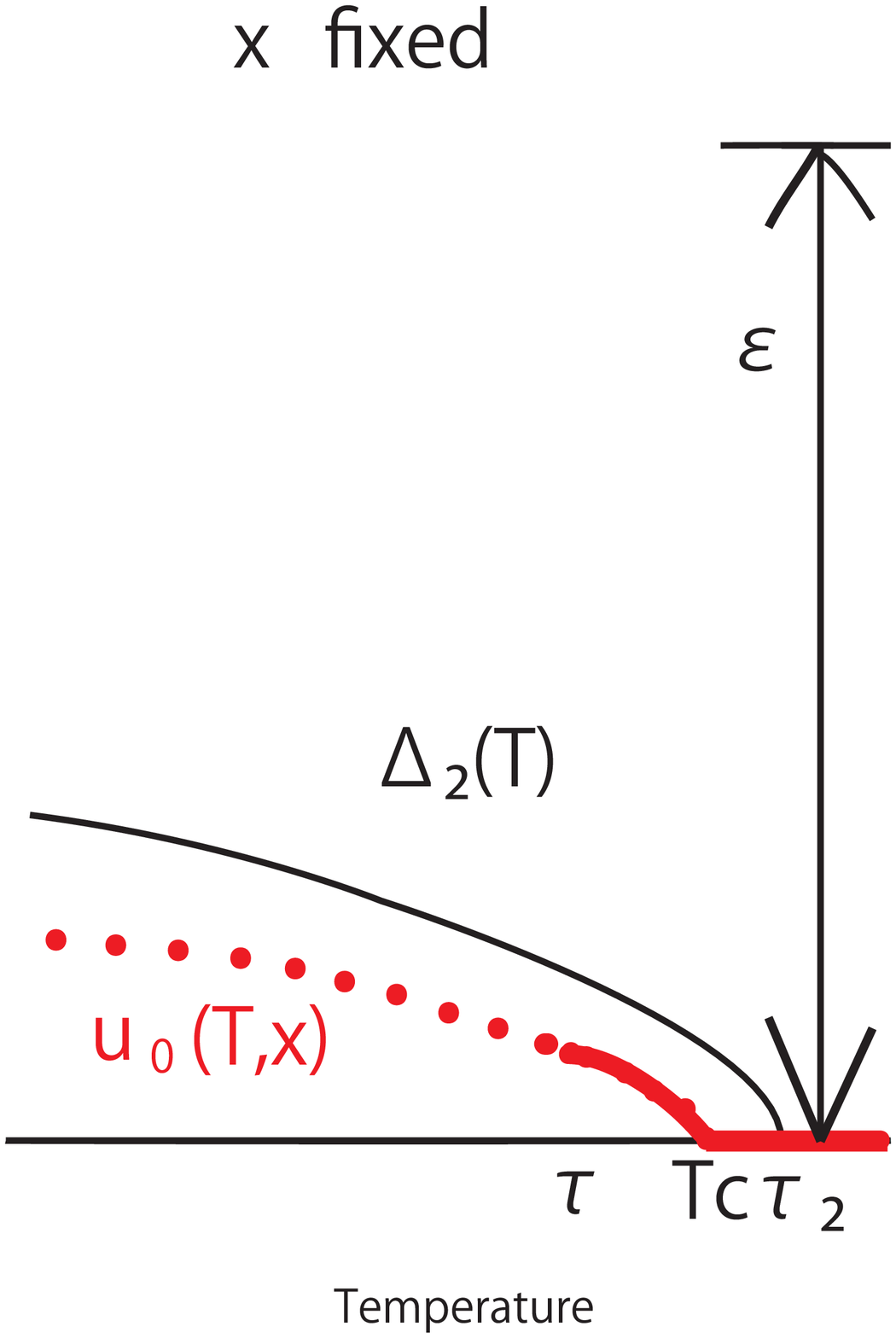}
\caption{\textsf{The graph of the solution $u_0 \in \overline{W}$ with the energy $x$ fixed.}}
\end{figure}
%%%%%%%%%%%%%%%%%%%%%%%%%%%

We then fix $\tau$ and $\varepsilon$ in \eqref{eqn:the-inequality}, and we deal with the set $[\tau,\, T_c] \times [\varepsilon,\,\hslash\omega_D] \in \mathbb{R}^2$. Note that the left side of \eqref{eqn:the-inequality} is a continuous function of $(T,\, x) \in [\tau,\, T_c] \times [\varepsilon,\,\hslash\omega_D]$. So we set
\begin{eqnarray*}
\alpha &=& \max_{(T,\, x) \in [\tau,\, T_c] \times [\varepsilon,\,\hslash\omega_D] }
\left[ \int_{\varepsilon}^{\hslash\omega_D} \frac{U(x,\,\xi)}{\,\sqrt{\,\xi^2+\Delta_2(T)^2\,}\,}\,
 \tanh \frac{\, \sqrt{\,\xi^2+\Delta_2(T)^2\,}\,}{2T}\,  d\xi \right. \\
&  & \left. \qquad \qquad \qquad \qquad \qquad +\frac{\, \Delta_2(\tau)^2 \,}{2\, \varepsilon^2}  \int_{\varepsilon}^{\hslash\omega_D}
  \frac{\, U(x,\,\xi) \,}{\xi}\, \tanh \frac{\xi}{\, 2T \,}\,  d\xi \right].  \nonumber
\end{eqnarray*}
Then
\begin{equation}\label{eq:alpha}
\alpha<1.
\end{equation}

We consider the following condition.

\noindent \textbf{Condition (C).} \     Let $\tau$ and $\varepsilon$ be as above. An element $u \in C([\tau,\, T_c] \times [\varepsilon,\,\hslash\omega_D])$ is partially differentiable with respect to the temperature $T \in [\tau,\, T_c)$ two times, and both $(\partial u/\partial T)$ and $(\partial^2 u/\partial T^2)$ belong to $C([\tau,\, T_c) \times [\varepsilon,\,\hslash\omega_D])$. Moreover, for the $u$ above, there are a unique $v \in C[\varepsilon,\,\hslash\omega_D]$ and a unique  $w \in C[\varepsilon,\,\hslash\omega_D]$ satisfying the following:

\noindent (C1) \   $v(x)>0$ at all $x \in [\varepsilon,\,\hslash\omega_D]$.\\
\noindent (C2) \   For an arbitrary $\varepsilon_1>0$, there is a $\delta>0$ such that $|T_c-T|<\delta$ implies
\[
\left| v(x)-\frac{\, u(T,\, x)^2 \,}{T_c-T} \right|<T_c\,\varepsilon_1 \quad \mbox{and} \quad \left| v(x)+2\, u(T,\, x)\, \frac{\, \partial u \,}{\partial T}(T,\, x) \right|<T_c\,\varepsilon_1\,.
\]
Here, the $\delta$ does not depend on $x \in [\varepsilon,\,\hslash\omega_D]$.\\
\noindent (C3) \   For an arbitrary $\varepsilon_1>0$, there is a $\delta>0$ such that $|T_c-T|<\delta$ implies
\[
\left| \frac{\, w(x)\,}{2}+\frac{\, u(T,\, x)^2+(T_c-T)\,\frac{\partial}{\,\partial T\,}\left\{u(T,\, x)^2\right\} \,}{(T_c-T)^2}  \right|<\varepsilon_1 \quad \mbox{and} \quad \left| w(x)-\frac{\partial^2}{\, \partial T^2\,}\left\{u(T,\, x)^2\right\} \right|<\varepsilon_1\,.
\]
Here, the $\delta$ does not depend on $x \in [\varepsilon,\,\hslash\omega_D]$.

\bigskip

We then define our operator $A$ (see \eqref{eqn:ouroperator}) on the following subset $W$ of the Banach space $C([\tau,\, T_c] \times [\varepsilon,\,\hslash\omega_D])$:
\[
Au(T,\,x)=\int_{\varepsilon}^{\hslash\omega_D}
\frac{U(x,\,\xi)\, u(T,\, \xi)}{\,\sqrt{\,\xi^2+u(T,\, \xi)^2\,}\,}\,
\tanh \frac{\,\sqrt{\,\xi^2+u(T,\, \xi)^2\,}\,}{2T}\, d\xi, \qquad u \in W,
\]
where
\begin{eqnarray}
W &=& \left\{ u \in C([\tau,\, T_c] \times [\varepsilon,\,\hslash\omega_D]) :  u(T,\,x) \geq u(T',\,x) \; \; (T<T'),
\right.  \nonumber \\
& & \left. \Delta_1(T) \leq u(T,\,x) \leq \Delta_2(T) \; \; \mbox{at} \; \; (T,\,x), \; (T',\,x) \in [\tau,\, T_c] \times [\varepsilon,\,\hslash\omega_D], \right. \nonumber \\
& & \left. u \; \mbox{satisfies Condition (C) above} \right\}. \nonumber
\end{eqnarray}

\begin{remark}\label{rmk:wutc0}
It follows directly from Condition (C2) that $u(T_c \,,\, x)=0$ at all $x \in [\varepsilon,\,\hslash\omega_D]$ for $u \in W$. 
\end{remark}

We denote by $\overline{W}$ the closure of the subset $W$ with respect to the norm $\| \cdot \|$ of the Banach space $C([\tau,\, T_c] \times [\varepsilon,\,\hslash\omega_D])$. 

\smallskip

The following are our main results.

\begin{theorem}\label{thm:solution}
Let $U(\cdot,\,\cdot)$ satisfy \eqref{eqn:conditionUprime}. Choose $\tau>0$ and $\varepsilon>0$ such that \eqref{eq:alpha} holds true. Then the operator $A: \overline{W} \to \overline{W}$ is a contraction operator, and hence there is a unique fixed point $u_0 \in \overline{W}$ of the operator $A: \overline{W} \to \overline{W}$. Consequently, there is a unique nonnegative solution $u_0 \in \overline{W}$ to the BCS-Bogoliubov gap equation \eqref{eqn:bcsgapeq}:
\[
u_0(T,\,x)=\int_{\varepsilon}^{\hslash\omega_D}
\frac{U(x,\,\xi)\, u_0(T,\, \xi)}{\,\sqrt{\,\xi^2+u_0(T,\, \xi)^2\,}\,}\,
\tanh \frac{\,\sqrt{\,\xi^2+u_0(T,\, \xi)^2\,}\,}{2T}\, d\xi, \quad (T,\, x) \in [\tau, \, T_c] \times [\varepsilon, \, \hslash\omega_D] \,.
\]
The solution $u_0$ is continuous on $[\tau,\, T_c] \times [\varepsilon,\,\hslash\omega_D]$, and is monotone decreasing with respect to the temperature $T$. Moreover, the solution $u_0$ satisfies that $\displaystyle{ \Delta_1(T) \leq u(T,\,x) \leq \Delta_2(T) }$ at all $(T,\, x) \in [\tau,\, T_c] \times [\varepsilon,\,\hslash\omega_D]$ and that $u_0(T_c \,,\, x)=0$ at all $x \in [\varepsilon,\,\hslash\omega_D]$. If $u_0 \in W$, then the solution $u_0$ satisfies Condition (C). On the other hand, if $u_0 \in \overline{W} \setminus W$, then the solution $u_0$ is approximated by an element of the subset $W$ fulfilling Condition (C).
\end{theorem}

See figure 2 for the graph of the solution $u_0 \in \overline{W}$ with the energy $x$ fixed. Since $u_0 \in \overline{W}$ by Theorem \ref{thm:solution}, we have $u_0 \in W$ or $u_0 \in \overline{W} \setminus W$. If  $u_0 \in \overline{W} \setminus W$, then the solution $u_0$ is approximated by a suitably chosen element $u_1 \in W$, as mentioned in Theorem \ref{thm:solution}. In \eqref{eqn:thermopotential} we then replace the solution $u_0 \in \overline{W} \setminus W$ by this  element $u_1 \in W$. Once we replace the solution $u_0 \in \overline{W} \setminus W$ of \eqref{eqn:thermopotential} by this $u_1 \in W$, we see that all the conditions of Definition \ref{dfn:thermodynamic} are satisfied. We immediately have the following.

\begin{theorem}\label{thm:2ndorderphase}

\noindent {\rm (1)} \   Suppose that $u_0 \in W$. Then all the conditions of Definition \ref{dfn:thermodynamic} are satisfied. Consequently the transition from a normal conducting state to a superconducting state at $T=T_c$ is a second-order phase transition.

\noindent {\rm (2)} \   Suppose that $u_0 \in \overline{W} \setminus W$. In \eqref{eqn:thermopotential}, we replace $u_0 \in \overline{W} \setminus W$ by a suitably chosen element $u_1 \in W$. Then all the conditions of Definition \ref{dfn:thermodynamic} are satisfied. Consequently the transition from a normal conducting state to a superconducting state at $T=T_c$ is a second-order phase transition.
\end{theorem}

Let $g\, : \, [0, \, \infty) \to \mathbb{R}$ be given by
\begin{equation}\label{eqn:function-g}
g(\eta)= \begin{cases}
     \displaystyle{ \frac{1}{\,\eta^2 \cosh^2\eta \,}-\frac{\,\tanh \eta \,}{\eta^3} } & \text{$(\eta>0)$}, \\
     \, \\
     \displaystyle{ -\frac{2}{\, 3 \,}  }   &  \text{$(\eta=0)$}.
            \end{cases}
\end{equation}
Note that $g(\eta)<0$. See Lemma \ref{lm:function-g} for some properties of the function $g$.

We remind here that the gap $\Delta C_V$ in the specific heat at constant volume at $T=T_c$ is given by Remark \ref{rmk:gapcv}. The following gives the exact and explicit expression for the gap.
\begin{proposition}\label{prp:gapheat}
Let $v$ be as in \rm{(C2)} of Condition \rm{(C)}, and $g$ as above. Then the gap $\Delta C_V$ in the specific heat at constant volume at $T=T_c$ is given by
\[
\Delta C_V=-\frac{N_0}{\, 8\, T_c\,} \, \int_{\varepsilon/(2T_c)}^{\hslash\omega_D/(2T_c)}
 v(2\, T_c\, \eta)^2 g(\eta) \, d\eta \quad (>0).
\]
\end{proposition}

%%%%%%%%%%%%%%%%%%%%%%%%%%%%%%%%%%%%%%%%%%%%%%%%%%%%%%%%%%%%%%% 3
\section{Proof of Theorem \ref{thm:solution}}

In this section we give a proof of Theorem \ref{thm:solution}. We first show that $A\, : \, W \to W$.

\begin{lemma}
If $u \in W$, then $Au \in C([\tau,\, T_c] \times [\varepsilon,\,\hslash\omega_D])$. 
\end{lemma}

\begin{proof}
Let $(T,\,x), \, (T_1,\,x_1) \in [\tau,\, T_c] \times [\varepsilon,\,\hslash\omega_D]$. For $u \in W$,
\begin{equation}\label{eq:Auconti}
Au(T,\,x)-Au(T_1,\,x_1)=Au(T,\,x)-Au(T,\,x_1)+Au(T,\,x_1)-Au(T_1,\,x_1).
\end{equation}

By \eqref{eqn:conditionUprime} the potential $U(\cdot,\,\cdot)$ is uniformly continuous on $ [\varepsilon,\,\hslash\omega_D]^2$, and hence for an arbitrary $\varepsilon_1>0$, there is a $\delta_1>0$ such that $|x-x_1|<\delta_1$ implies
\[
\left| U(x,\,\xi)-U(x_1,\,\xi) \right|<\frac{\varepsilon_1}{\,3\hslash\omega_D\,} \quad \mbox{at all} \quad \xi \in [\varepsilon,\,\hslash\omega_D]. 
\]
Note that the $\delta_1$ does not depend nor on $x$, nor on $x_1$, nor on $\xi$, nor on $T$, nor on $T_1$. The first and second terms on the right side of \eqref{eq:Auconti} therefore turn into
\begin{eqnarray*}
\left| Au(T,\,x)-Au(T,\,x_1) \right|
&\leq& \int_{\varepsilon}^{\hslash\omega_D}
\frac{\left| U(x,\,\xi)-U(x_1\,,\,\xi) \right|\, u(T,\, \xi)}{\,\sqrt{\,\xi^2+u(T,\, \xi)^2\,}\,}\,
\tanh \frac{\,\sqrt{\,\xi^2+u(T,\, \xi)^2\,}\,}{2T}\, d\xi \nonumber \\
&\leq& \int_{\varepsilon}^{\hslash\omega_D}
\left| U(x,\,\xi)-U(x_1\,,\,\xi) \right| \, d\xi \nonumber \\
&<& \frac{\,\varepsilon_1\,}{3}\, .\nonumber \\
\end{eqnarray*}
On the other hand, the third and fourth terms become
\begin{equation}\label{eqn:monotone-decreasing}
Au(T,\,x_1)-Au(T_1,\,x_1) = \int_{\varepsilon}^{\hslash\omega_D} U(x_1\,,\,\xi)
\left\{ K_1+K_2 \right\}\, d\xi,
\end{equation}
where
\begin{eqnarray*}
K_1 &=& \frac{u(T,\, \xi)}{\,\sqrt{\,\xi^2+u(T,\, \xi)^2\,}\,}\,
\tanh \frac{\,\sqrt{\,\xi^2+u(T,\, \xi)^2\,}\,}{2T} \nonumber \\
& & \qquad -\frac{u(T_1\,,\, \xi)}{\,\sqrt{\,\xi^2+u(T_1\,,\, \xi)^2\,}\,}\,
\tanh \frac{\,\sqrt{\,\xi^2+u(T_1\,,\, \xi)^2\,}\,}{2T}, \nonumber \\
K_2 &=& \frac{u(T_1\,,\, \xi)}{\,\sqrt{\,\xi^2+u(T_1\,,\, \xi)^2\,}\,}
\left\{ \tanh \frac{\,\sqrt{\,\xi^2+u(T_1\,,\, \xi)^2\,}\,}{2T}-\tanh \frac{\,\sqrt{\,\xi^2+u(T_1\,,\, \xi)^2\,}\,}{2T_1} \right\}. \nonumber \\
\end{eqnarray*}
Note that $u \in W$ is uniformly continuous on $[\tau,\, T_c] \times [\varepsilon,\,\hslash\omega_D]$. Then, for the $\varepsilon_1>0$ above, there is a $\delta_2>0$ such that $| T-T_1|<\delta_2$ implies
\[
\left| u(T,\,\xi)-u(T_1\,,\,\xi)  \right|<\frac{\varepsilon_1}{\, 3\, \alpha \,}.
\]
Here, $\alpha$ is that in \eqref{eq:alpha}, and the $\delta_2$ does not depend nor on $x$, nor on $x_1$, nor on $\xi$, nor on $T$, nor on $T_1$. However, the $\delta_2$ may depend on $u \in W$. Note that $\displaystyle{ \frac{z}{\, \cosh^2z\,} \leq \tanh z }$ \   $(z \geq 0)$. Hence
\begin{eqnarray*}
& & \int_{\varepsilon}^{\hslash\omega_D} U(x_1\,,\,\xi) \,  \left| K_1 \right| \, d\xi, \nonumber \\
&=& \int_{\varepsilon}^{\hslash\omega_D} 
     \frac{U(x_1\,,\,\xi)}{\, (\xi^2+c^2)^{3/2} \,}\left\{ \xi^2\tanh \frac{\,\sqrt{\,\xi^2+c^2\,}\,}{2T}
     +c^2\frac{\, \sqrt{\, \xi^2+c^2\,}\,}{2T}\frac{1}{\, \cosh^2 \frac{\, \sqrt{\,\xi^2+c^2 \,}\,}{2T} \,} \right\}
     \nonumber \\
& & \qquad \times \left| u(T,\,\xi)-u(T_1\,,\,\xi) \right| \, d\xi \nonumber \\
&\leq& \int_{\varepsilon}^{\hslash\omega_D} 
      \frac{\, U(x_1\,,\,\xi)\,}{\, \sqrt{ \, \xi^2+c^2 \, } \,} \tanh \frac{\, \sqrt{ \, \xi^2+c^2 \, } \,}{2T} 
      \, \left| u(T,\,\xi)-u(T_1\,,\,\xi) \right| \, d\xi.
\end{eqnarray*}
Here, $c$ is between $u(T,\,\xi)$ and $u(T_1\,,\,\xi)$. Note again that $\displaystyle{ \frac{z}{\, \cosh^2z\,} \leq \tanh z }$ \   $(z \geq 0)$ and that the function $\displaystyle{ z \mapsto \frac{\,\tanh z\,}{z} }$ \   $(z \geq 0)$ is strictly decreasing. Then a straightforward calculation gives
\begin{eqnarray}\label{eqn:keyinequality}
& &\frac{1}{\, \sqrt{ \, \xi^2+c^2 \, } \,} \tanh \frac{\, \sqrt{ \, \xi^2+c^2 \, } \,}{2T} \\
&=&
\frac{1}{\, \sqrt{ \, \xi^2+\Delta_2(T)^2 \, } \,} \tanh \frac{\, \sqrt{ \, \xi^2+\Delta_2(T)^2 \, } \,}{2T} 
    \nonumber \\
& & \qquad +\frac{\Delta_2(T)^2-c^2}{\, 2(\xi^2+c_1^2)^{3/2} \,}
 \left\{ \tanh \frac{\, \sqrt{ \, \xi^2+c_1^2 \, } \,}{2T} 
  -\frac{\, \sqrt{ \, \xi^2+c_1^2 \, } \,}{2T} \frac{1}{\, \cosh^2 \frac{\, \sqrt{\,\xi^2+c_1^2 \,}\,}{2T} \,}
 \right\} \nonumber \\
&\leq&
\frac{1}{\, \sqrt{ \, \xi^2+\Delta_2(T)^2 \, } \,} \tanh \frac{\, \sqrt{ \, \xi^2+\Delta_2(T)^2 \, } \,}{2T}
+\frac{\, \Delta_2(\tau)^2 \,}{2\, \varepsilon^2} \frac{1}{\, \xi \, } \tanh \frac{\xi}{\, 2T \,}, \nonumber
\end{eqnarray}
where $c_1$ satisfies $c<c_1<\Delta_2(T)$. Hence
\[
\int_{\varepsilon}^{\hslash\omega_D} U(x_1\,,\,\xi) \,  \left| K_1 \right| \, d\xi <\frac{\varepsilon_1}{\, 3 \, \alpha \,} \, \alpha=\frac{\varepsilon_1}{\, 3 \,}.
\]
Moreover, if $\displaystyle{ |T-T_1|<\frac{2\, \tau^2\varepsilon_1}{\, 3\, U_2\, \Delta_2(\tau)\, \hslash\omega_D \,} }$, then
\begin{eqnarray*}
\int_{\varepsilon}^{\hslash\omega_D} U(x_1\,,\,\xi) \, \left| K_2 \right| \, d\xi
&\leq& U_2\int_{\varepsilon}^{\hslash\omega_D} 
     \frac{u(T_1\,,\,\xi)}{\, 2(T'')^2 \cosh^2 \frac{\, \sqrt{\,\xi^2+u(T_1,\,\,\xi)^2 \,}\,}{2T''} \,}\, d\xi \;
     |T-T_1| \nonumber \\
&\leq& U_2 \int_{\varepsilon}^{\hslash\omega_D} \frac{\,\Delta_2(\tau)\,}{2\, \tau^2}\, d\xi \;
     |T-T_1| \nonumber \\
&<& \frac{\,\varepsilon_1\,}{3} \,.
\end{eqnarray*}
Here, $T''$ is between $T$ and $T_1$. Thus
\[
| Au(T,\,x)-Au(T_1,\,x_1)| < \varepsilon_1\,, 
\]
where $\displaystyle{ |T-T_1|+|x-x_1|<\delta=\min\left( \delta_1\,,\,\delta_2\,,\, \frac{2\, \tau^2\varepsilon_1}{\, 3\, U_2\,\Delta_2(\tau)\,\hslash\omega_D \,} \right)}$.
\end{proof}

\begin{lemma}
Let $(T,\,x), \, (T_1,\,x) \in [\tau,\, T_c] \times [\varepsilon,\,\hslash\omega_D]$, and let $T<T_1$.
If $u \in W$, then $Au(T,\,x) \geq Au(T_1,\,x)$.
\end{lemma}

\begin{proof}
Since $T<T_1$,
\[
\frac{u(T,\, \xi)}{\,\sqrt{\,\xi^2+u(T,\, \xi)^2\,}\,} \geq
\frac{u(T_1\,,\, \xi)}{\,\sqrt{\,\xi^2+u(T_1\,,\, \xi)^2\,}\,}.
\]
Hence, $K_1 \geq 0$ and $K_2 \geq 0$ in \eqref{eqn:monotone-decreasing}. Thus
\[
Au(T,\,x)-Au(T_1,\,x) \geq 0.
\]
\end{proof}

\begin{lemma}
Let $(T,\, x) \in [\tau,\, T_c] \times [\varepsilon,\,\hslash\omega_D]$.
If $u \in W$, then $\Delta_1(T) \leq Au(T,\,x) \leq \Delta_2(T)$.
\end{lemma}

\begin{proof}
Since
\[
 \frac{u(T,\, \xi)}{\,\sqrt{\,\xi^2+u(T,\, \xi)^2\,}\,} \leq \frac{\Delta_2(T)}{\,\sqrt{\,\xi^2+\Delta_2(T)^2\,}\,}\, ,
\]
it follows from \eqref{eqn:delta2} that
\[
Au(T,\, x) \leq U_2\int_{\varepsilon}^{\hslash\omega_D}
 \frac{\Delta_2(T)}{\,\sqrt{\,\xi^2+\Delta_2(T)^2\,}\,}\,
 \tanh \frac{\, \sqrt{\,\xi^2+\Delta_2(T)^2\,}\,}{2T}\,d\xi=\Delta_2(T).
\]
Similarly we can show that $\Delta_1(T) \leq Au(T,\, x)$.
\end{proof}

We now show that $Au$ \    ($u \in W$) satisfies Condition (C) so as to conclude that $A\, : \, W \to W$. A straightforward calculation gives the following.
\begin{lemma}
Let $u \in W$. Then $Au$ is partially differentiable with respect to $T \in [\tau,\, T_c)$ twice, and
\[
\frac{\, \partial Au \,}{\partial T}, \;  \frac{\, \partial^2 Au \,}{\partial T^2} \in C([\tau,\, T_c) \times [\varepsilon,\,\hslash\omega_D]).
\]   
\end{lemma}

For $u \in W$, let $v$ be as in Condition (C). Note that $v$ depends on the $u$. Set
\begin{equation}\label{eqn:functionF}
F(x)=\left\{ \int_{\varepsilon}^{\hslash\omega_D} U(x,\,\xi)
\frac{\, \sqrt{ v(\xi) }\, }{\xi} \, \tanh \frac{\xi}{\, 2T_c \, }\, d\xi \right\}^2 \quad (>0), \qquad \varepsilon \leq x \leq \hslash\omega_D \,. 
\end{equation}

\begin{lemma}\label{lm:vandF}
Suppose $u \in W$. Then the function $F$ given by \eqref{eqn:functionF} belongs to $C[\varepsilon,\,\hslash\omega_D]$, and for an arbitrary $\varepsilon_1>0$, there is a $\delta>0$ such that $|T_c-T|<\delta$ implies
\[
\left| F(x)-\frac{\, \{ Au(T,\, x) \}^2 \,}{T_c-T} \right|<T_c\,\varepsilon_1 \quad \mbox{and} \quad \left| F(x)+2\, Au(T,\, x)\, \frac{\, \partial Au \,}{\partial T}(T,\, x) \right|<T_c\,\varepsilon_1.
\]
Here, the $\delta$ does not depend on $x \in [\varepsilon,\,\hslash\omega_D]$. Such a function $F$ is uniquely given by \eqref{eqn:functionF}.
\end{lemma}

\begin{proof}
Since the potential $U(\cdot,\,\cdot)$ is uniformly continuous on $ [\varepsilon,\,\hslash\omega_D]^2$ by \eqref{eqn:conditionUprime}, the function $F$ is continuous on $[\varepsilon,\,\hslash\omega_D]$. Moreover,
\[
\left| F(x)-\frac{\, \{ Au(T,\, x) \}^2 \,}{T_c-T} \right| \leq U_2^2 \, I_0 \, \left( I_1+I_2+I_3 \right),
\]
where
\begin{eqnarray*}
I_0 &=& \int_{\varepsilon}^{\hslash\omega_D}  \frac{1}{\, \xi \,}
 \left( \sqrt{\, v(\xi)\,}+\sqrt{\, \frac{\, u(T,\,\xi)^2 \,}{T_c-T} \,} \right) \, d\xi, \nonumber \\
I_1 &=& \int_{\varepsilon}^{\hslash\omega_D} \frac{1}{\, \xi \,}
 \left| \sqrt{\, v(\xi)\,}-\sqrt{\, \frac{\, u(T,\,\xi)^2 \,}{T_c-T} \,} \right| \tanh\frac{\xi}{\,2T_c\,} \, d\xi,                                       \nonumber \\
I_2 &=& \int_{\varepsilon}^{\hslash\omega_D} \sqrt{\, \frac{\, u(T,\,\xi)^2 \,}{T_c-T} \,}
  \left|  \frac{1}{\, \xi \,}-\frac{1}{\sqrt{\, \xi^2+u(T,\,\xi)^2 \,}} \right| \tanh\frac{\xi}{\,2T_c\,} \, d\xi, \nonumber \\
I_3 &=& \int_{\varepsilon}^{\hslash\omega_D}
  \sqrt{\, \frac{\, u(T,\,\xi)^2 \,}{T_c-T} \,} \frac{1}{\sqrt{\, \xi^2+u(T,\,\xi)^2 \,}}
  \left| \tanh\frac{\xi}{\,2T_c\,}-\tanh\frac{\, \sqrt{\, \xi^2+u(T,\,\xi)^2 \,} \,}{\,2T\,}  \right| \, d\xi.
\nonumber \\
\end{eqnarray*}
By Condition (C2), for $(0<)\varepsilon_1<1$, there is a $\delta_1>0$ such that $T_c-T<\delta_1$ implies
\[
\frac{\, u(T,\, \xi)^2 \,}{T_c-T}<v(\xi)+T_c\,\varepsilon_1<v(\xi)+T_c\,.
\]
Note that the $\delta_1$ does not depend nor on $x$, nor on $\xi$. Moreover, for $(0<)\varepsilon_1<1$, there is a $\delta_2>0$ such that $T_c-T<\delta_2$ implies
\[
u(T,\, \xi)^2=\frac{\, u(T,\, \xi)^2 \,}{T_c-T}\, (T_c-T)
<\left\{ \max_{\xi \in [\varepsilon,\,\hslash\omega_D]}v(\xi)+T_c \right\} \,(T_c-T)
<T_c^2 \, \varepsilon_1^2 \,.
\]
Here, $\delta_2=\displaystyle{
\frac{T_c^2\varepsilon_1^2}{\, \max_{\xi \in [\varepsilon,\,\hslash\omega_D]}v(\xi)+T_c \,} }$ \   and the $\delta_2$ does not depend nor on $x$, nor on $\xi$.

Noting $\displaystyle{ \frac{\,\tanh z\,}{z}\leq 1 }$ \quad  $(z\geq 0)$, we find
\begin{eqnarray*}
I_0 &<& 2\sqrt{\, \displaystyle{ \max_{\xi \in [\varepsilon,\,\hslash\omega_D]}v(\xi)+T_c  } \,} \;
\ln \frac{\, \hslash\omega_D \,}{\varepsilon}\,, \nonumber \\
I_1 &<& \frac{\, \varepsilon_1 \,}{2} \int_{\varepsilon}^{\hslash\omega_D}
 \frac{d\xi}{\,\sqrt{\, v(\xi)\,} \,},  \nonumber \\
I_2 &\leq& \sqrt{\, \displaystyle{ \max_{\xi \in [\varepsilon,\,\hslash\omega_D]}v(\xi)+T_c  } \,} \;
 \int_{\varepsilon}^{\hslash\omega_D} \frac{\, u(T,\, \xi) \,}{\, 2T_c\,\xi \,} \,d\xi
 < \sqrt{\, \displaystyle{ \max_{\xi \in [\varepsilon,\,\hslash\omega_D]}v(\xi)+T_c  } \,} \;
  \frac{\, \varepsilon_1 \,}{2}\, \ln \frac{\, \hslash\omega_D \,}{\varepsilon}\,, \nonumber \\
I_3 &\leq& \sqrt{\, \displaystyle{ \max_{\xi \in [\varepsilon,\,\hslash\omega_D]}v(\xi)+T_c  } \,} \;
 \int_{\varepsilon}^{\hslash\omega_D} \left\{ 
 \frac{\, u(T,\, \xi) \,}{\,4\tau \xi \,}+\frac{\, T_c-T \,}{2\tau^2}   \right\} \,d\xi \nonumber \\
 &<& \sqrt{\, \displaystyle{ \max_{\xi \in [\varepsilon,\,\hslash\omega_D]}v(\xi)+T_c  } \,} \;
  \left( \frac{1}{\, 4\tau \,}\, \ln \frac{\, \hslash\omega_D \,}{\varepsilon}
          +\frac{\, \hslash\omega_D \,}{2\tau^2}  \right)T_c \, \varepsilon_1 \,. \nonumber \\
\end{eqnarray*}
Here, $T_c-T<\delta=\min(\delta_1\,,\, \delta_2\,,\, T_c \, \varepsilon_1)$. Note that the $\delta$ does not depend on $x \in [\varepsilon,\,\hslash\omega_D]$. Uniqueness of $F$ follows immediately.

We can show
\[
\left| F(x)+2\, Au(T,\, x)\, \frac{\, \partial Au \,}{\partial T}(T,\, x) \right|<T_c\,\varepsilon_1
\]
similarly.
\end{proof}

For $u \in W$, let $v$ and $w$ be as in Condition (C). Note that $v$ and $w$ depend on the $u$. Set
\begin{eqnarray}\label{eqn:functionG}
& & G(x)  \nonumber \\
&=& \int_{\varepsilon}^{\hslash\omega_D} U(x,\,\xi)
 \frac{\, \sqrt{ v(\xi) }\, }{\xi} \, \tanh \frac{\xi}{\, 2T_c \, }\, d\xi \times \nonumber \\ \nonumber
& & \int_{\varepsilon}^{\hslash\omega_D} U(x,\,\eta) \left\{
 \left( \frac{\, w(\eta)\, }{\,\eta\sqrt{\, v(\eta)\,}\,}-\frac{\, 2\sqrt{\,v(\eta)^3  \,} \,}{\eta^3} \right) 
 \, \tanh \frac{\eta}{\, 2T_c \, }+
 \frac{\sqrt{\, v(\eta)\,}}{\, \cosh^2\frac{\eta}{\, 2T_c\,} \,}\left(
 \frac{v(\eta)}{\, \eta^2T_c \,}+\frac{2}{\, T_c^2 \,} \right)
 \right\} \, d\eta, \\
\end{eqnarray}
where $\varepsilon \leq x \leq \hslash\omega_D$.

\begin{lemma}
Suppose $u \in W$. Then the function $G$ given by \eqref{eqn:functionG} belongs to $C[\varepsilon,\,\hslash\omega_D]$, and for an arbitrary $\varepsilon_1>0$, there is a $\delta>0$ such that $|T_c-T|<\delta$ implies
\[
\left| \frac{\, G(x)\,}{2}+\frac{\, \{Au(T,\, x)\}^2+(T_c-T)\,\frac{\partial}{\,\partial T\,}\left\{ Au(T,\, x)^2\right\} \,}{(T_c-T)^2} \right|<\varepsilon_1
\]
and
\[
\left| G(x)-\frac{\partial^2}{\, \partial T^2\,}\left\{ Au(T,\, x)^2 \right\} \right|<\varepsilon_1\,.
\]
Here, the $\delta$ does not depend on $x \in [\varepsilon,\,\hslash\omega_D]$. Such a function $G$ is uniquely given by \eqref{eqn:functionG}.
\end{lemma}

\begin{proof}
Since the potential $U(\cdot,\,\cdot)$ is uniformly continuous on $ [\varepsilon,\,\hslash\omega_D]^2$ by \eqref{eqn:conditionUprime}, the function $G$ is continuous on $[\varepsilon,\,\hslash\omega_D]$. A discussion similar to that in the proof of Lemma \ref{lm:vandF} shows the rest. This time we also need Condition (C3) in Section 2.
\end{proof}

The lemmas above immediately give the following:
\begin{lemma}
$\displaystyle{\quad A\, : \, W \to W. }$
\end{lemma}

As mentioned above, we denote by $\| \cdot \|$ the norm of the Banach space $C([\tau,\, T_c] \times [\varepsilon,\,\hslash\omega_D])$.

\begin{lemma}\label{lm:estimate}
Let $\alpha$ be as in \eqref{eq:alpha}. Then \    $\left\| Au-Av \right\| \leq \alpha \| u-v \|$ for $u,\,v \in W$.
\end{lemma}

\begin{proof}
Let $u,\,v \in W$. Let $c$ be between $u(T,\, \xi)$ and $v(T,\, \xi)$. Since $\displaystyle{ \frac{z}{\,\cosh^2z\,} \leq \tanh z }$ \   $(z \geq 0)$, it then follows that
\begin{eqnarray*}
& & \left|  Au(T,\,x)-Av(T,\, x) \right| \nonumber \\
&\leq& \int_{\varepsilon}^{\hslash\omega_D} U(x,\,\xi) \left|
 \frac{u(T,\, \xi)}{\,\sqrt{\,\xi^2+u(T,\, \xi)^2\,}\,}\, \tanh \frac{\,\sqrt{\,\xi^2+u(T,\, \xi)^2\,}\,}{2T}
\right. \nonumber \\
& & \qquad \left.  -\frac{v(T,\, \xi)}{\,\sqrt{\,\xi^2+v(T,\, \xi)^2\,}\,}\,
 \tanh \frac{\,\sqrt{\,\xi^2+v(T,\, \xi)^2\,}\,}{2T}  \right| \, d\xi \nonumber \\
&\leq& \int_{\varepsilon}^{\hslash\omega_D} \frac{\, U(x,\,\xi)\,}{\, (\xi^2+c^2)^{3/2}\,} \left\{
 \xi^2 \tanh \frac{\,\sqrt{\,\xi^2+c^2\,}\,}{2T}
 +c^2\frac{\,\sqrt{\,\xi^2+c^2\,}\,}{2T}\frac{1}{\, \cosh^2 \frac{\,\sqrt{\,\xi^2+c^2\,}\,}{2T} \,}  \right\}\, d\xi \;
\| u-v \| \nonumber \\
&\leq& \int_{\varepsilon}^{\hslash\omega_D} \frac{\, U(x,\,\xi)\,}{\, \sqrt{ \, \xi^2+c^2 \,}\,} \,
 \tanh \frac{\,\sqrt{\,\xi^2+c^2\,}\,}{2T} \, d\xi \;  \| u-v \|. \nonumber
\end{eqnarray*}
A discussion similar to that in \eqref{eqn:keyinequality} gives
\begin{eqnarray*}
& & \left|  Au(T,\,x)-Av(T,\, x) \right| \\
&\leq& \int_{\varepsilon}^{\hslash\omega_D} U(x,\,\xi) \,\left\{
\frac{1}{\, \sqrt{ \, \xi^2+\Delta_2(T)^2 \, } \,} \tanh \frac{\, \sqrt{ \, \xi^2+\Delta_2(T)^2 \, } \,}{2T}
+\frac{\, \Delta_2(\tau)^2 \,}{2\, \varepsilon^2} \frac{1}{\, \xi \, } \tanh \frac{\xi}{\, 2T \,}
 \right\}  \, d\xi \times \\
& & \qquad \times \| u-v \| \\
&\leq& \alpha \,  \| u-v \| . 
\end{eqnarray*}
\end{proof}

We extend the domain $W$ of our operator $A$ to its closure $ \overline{W}$. Let $u \in \overline{W}$. Then there is a sequence $\{ u_n \}_{n=1}^{\infty} \subset W$ satisfying $\| u-u_n \| \to 0$ as $n \to \infty$. 
Lemma \ref{lm:estimate} gives $\{ Au_n \}_{n=1}^{\infty} \subset W$ is a Cauchy sequence, and hence there is an $Au \in \overline{W}$  satisfying $\| Au-Au_n \| \to 0$ as $n \to \infty$. Note that $Au \in \overline{W}$ does not depend on the sequence $\{ u_n \}_{n=1}^{\infty} \subset W$. We thus have the following.

\begin{lemma}
$A:\,  \overline{W} \to \overline{W}$.
\end{lemma}

It is not obvious that $Au$ \    $(u \in \overline{W})$ is expressed as that in \eqref{eqn:ouroperator}. The next lemma shows this is the case.

\begin{lemma}\label{lm:integrali}
Let $u \in \overline{W}$. Then
\[
Au(T,\,x)=\int_{\varepsilon}^{\hslash\omega_D}
\frac{U(x,\,\xi)\, u(T,\, \xi)}{\,\sqrt{\,\xi^2+u(T,\, \xi)^2\,}\,}\,
\tanh \frac{\,\sqrt{\,\xi^2+u(T,\, \xi)^2\,}\,}{2T}\, d\xi, \quad
(T,\,x) \in [\tau,\,T_c] \times [\varepsilon,\,\hslash\omega_D].
\]
\end{lemma}

\begin{proof}
For $u \in \overline{W}$, set
\[
I(T,\,x)=\int_{\varepsilon}^{\hslash\omega_D}
\frac{U(x,\,\xi)\, u(T,\, \xi)}{\,\sqrt{\,\xi^2+u(T,\, \xi)^2\,}\,}\,
\tanh \frac{\,\sqrt{\,\xi^2+u(T,\, \xi)^2\,}\,}{2T}\, d\xi
\]
and let $\{ u_n \}_{n=1}^{\infty} \subset W$ be a sequence satisfying $\| u-u_n \| \to 0$ as $n\to\infty$. Note that the function $(T,\,x) \mapsto I(T,\,x)$ is well-defined and continuous. Then
\[
| Au(T,\,x)-I(T,\,x) | \leq | Au(T,\,x)-Au_n(T,\,x) |+ | Au_n(T,\,x)-I(T,\,x) |.
\]
Since $\| Au-Au_n \| \to 0$ as $n\to\infty$, the first term on the right side becomes
\[
 | Au(T,\,x)-Au_n(T,\,x) | \leq  \| Au-Au_n \| \to 0 \quad  (n \to \infty).
\]
A discussion similar to that in the proof of Lemma \ref{lm:estimate} gives the second term becomes
\[
 | Au_n(T,\,x)-I(T,\,x) | \leq \alpha \left\| u_n-u \right\| \to 0 \quad  (n \to \infty).
\]
The result thus follows.
\end{proof}

Lemma \ref{lm:estimate} immediately gives the following.

\begin{lemma}\label{lm:estimate-prime}
Let $\alpha$ be as in \eqref{eq:alpha}. Then \    $\left\| Au-Av \right\| \leq a \| u-v \|$ for $u,\,v \in \overline{W}$. Consequently, the operator $A: \, \overline{W} \to \overline{W}$ is a contraction operator.
\end{lemma}

The Banach fixed-point theorem thus implies the following.

\begin{lemma}
The operator $A: \overline{W} \to \overline{W}$ has a unique fixed point $u_0 \in \overline{W}$. 
Consequently, there is a unique nonnegative solution $u_0 \in \overline{W}$ to the BCS-Bogoliubov gap equation \eqref{eqn:bcsgapeq}:
\[
u_0(T,\,x)=\int_{\varepsilon}^{\hslash\omega_D}
\frac{U(x,\,\xi)\, u_0(T,\, \xi)}{\,\sqrt{\,\xi^2+u_0(T,\, \xi)^2\,}\,}\,
\tanh \frac{\,\sqrt{\,\xi^2+u_0(T,\, \xi)^2\,}\,}{2T}\, d\xi, \quad (T,\, x) \in [\tau, \, T_c] \times [\varepsilon, \, \hslash\omega_D].
\]
\end{lemma}

Now our proof of Theorem \ref{thm:solution} is complete.

%%%%%%%%%%%%%%%%%%%%%%%%%%%%%%%%%%%%%%%%%%%%%%%%%%%%%%%%%%%%%%% 4
\section{Proofs of Theorem \ref{thm:2ndorderphase} and Proposition \ref{prp:gapheat}}

We begin this section by preparing a lemma. As mentioned in Theorem \ref{thm:2ndorderphase}, the function $u_0$ in \eqref{eqn:thermopotential} is the solution $u_0 \in \overline{W}$ of Theorem \ref{thm:solution}; however, if $u_0 \in \overline{W} \setminus W$, we then approximate $u_0 \in \overline{W} \setminus W$ by a suitably chosen element $u_1 \in W$ and we replace $u_0$ in \eqref{eqn:thermopotential} by this $u_1 \in W$. We denote by $\Psi_1(T)$ the thermodynamic potential corresponding to this element $u_1 \in W$:
\begin{eqnarray}\label{eqn:psi1}
\Psi_1(T)
&=& -2N_0 \int_{\varepsilon}^{\hslash\omega_D} \left\{ \sqrt{\,\xi^2+u_1(T,\, \xi)^2\,}-\xi \right\} \, d\xi \\
& & +N_0 \int_{\varepsilon}^{\hslash\omega_D} \frac{u_1(T,\, \xi)^2}{\,\sqrt{\,\xi^2+u_1(T,\, \xi)^2\,}\,}\,
\tanh \frac{\,\sqrt{\,\xi^2+u_1(T,\, \xi)^2\,}\,}{2T}\, d\xi \nonumber \\
& & -4N_0 T \int_{\varepsilon}^{\hslash\omega_D} \ln
\frac{\, 1+e^{ -\sqrt{\,\xi^2+u_1(T,\, \xi)^2\,}/T } \,}{  1+e^{-\xi/T}  } \, d\xi, \quad T \in [\tau, \, T_c].
 \nonumber \\ \nonumber
\end{eqnarray}

A discussion similar to that in the proof of Lemma \ref{lm:estimate} gives the following, which shows that $\Psi$ is approximated by $\Psi_1$.
\begin{lemma}\label{lm:prepare}
Let $\Psi$ be as in \eqref{eqn:thermopotential} and $\Psi_1$ as in \eqref{eqn:psi1}. Then, at all $T \in [\tau,\, T_c]$,
\[
\left| \Psi(T)-\Psi_1(T) \right| \leq 2N_0\,\Delta_2(0)\left\{
 \left( 1+2\frac{\, T_c\,}{\tau} \right)\, \ln\frac{\, \hslash\omega_D\,}{\varepsilon}+\alpha \right\}
 \| u-u_0 \|,
\]
where $\alpha$ is that in \eqref{eq:alpha}.
\end{lemma}

\begin{remark}\label{rmk:replacement}
In what follows, when the solution $u_0$ to the BCS-Bogoliubov gap equation \eqref{eqn:bcsgapeq} is an element of $W$, we denote by $u$ below the very solution $u_0 \in W$; when the solution $u_0$ is an element of $\overline{W} \setminus W$, we denote by $u$ below the suitably chosen element $u_1 \in W$ mentioned just above. Therefore, in what follows, the function $u$ does not always denote the solution and is an element of $W$.
\end{remark}

\begin{lemma}\label{lm:psia}
Let $\Psi$ be as in \eqref{eqn:thermopotential}. Then $\Psi$ is differentiable on $[\tau,\, T_c]$, and
\[
\Psi(T_c)=0 \quad \mbox{and} \quad \frac{\, \partial \Psi \,}{\partial T}(T_c)=0.
\]
\end{lemma}

\begin{proof} \    
Note that $u \in W$, as mentioned in Remark \ref{rmk:replacement}. It then follows that $u(T_c \,,\, \xi)=0$ at all $\xi \in [\varepsilon,\,\hslash\omega_D]$ (see Remark \ref{rmk:wutc0} above). Hence $\Psi(T_c)=0$. A straightforward calculation gives that $\Psi$ is differentiable on $[\tau,\, T_c)$. So it suffices to show that $\Psi$ is differentiable at $T=T_c$ and that $(\partial \Psi/\partial T)(T_c)=0$. Note that $\Psi(T_c)=0$. Then
\begin{eqnarray}\label{eqn:psi3}
\quad  \frac{\, \Psi(T_c)-\Psi(T) \,}{T_c-T}
&=& 2N_0 \int_{\varepsilon}^{\hslash\omega_D}
 \frac{\, u(T,\, \xi)^2 \,}{T_c-T} \, \frac{1}{\,\sqrt{\,\xi^2+u(T,\, \xi)^2\,}+\xi\,} \, d\xi \\
& & -N_0 \int_{\varepsilon}^{\hslash\omega_D}
 \frac{\, u(T,\, \xi)^2 \,}{T_c-T} \, \frac{1}{\,\sqrt{\,\xi^2+u(T,\, \xi)^2\,}\,} \,
\tanh \frac{\,\sqrt{\,\xi^2+u(T,\, \xi)^2\,}\,}{2T} \, d\xi \nonumber \\
& & +4N_0 T \int_{\varepsilon}^{\hslash\omega_D} \frac{1}{\, T_c-T \,}
 \ln \frac{\, 1+e^{ -\sqrt{\,\xi^2+u(T,\, \xi)^2\,}/T } \,}{  1+e^{-\xi/T}  } \, d\xi. \nonumber \\ \nonumber
\end{eqnarray}
By (C2) of Condition (C), for an arbitrary $(0<) \varepsilon_1<1$, there is a $\delta>0$ such that $| T_c-T |<\delta$ implies
\[
\frac{\, u(T,\, \xi)^2 \,}{T_c-T} \, \frac{1}{\,\sqrt{\,\xi^2+u(T,\, \xi)^2\,}+\xi\,}
<\frac{\, v(\xi)+T_c\,\varepsilon_1\,}{\xi} < \frac{\, v(\xi)+T_c\,}{\xi}.
\]
The Lebesgue dominated convergence theorem therefore implies that the first term on the right side of \eqref{eqn:psi3} becomes
\[
2N_0 \lim_{T \uparrow T_c} \int_{\varepsilon}^{\hslash\omega_D}
 \frac{\, u(T,\, \xi)^2 \,}{T_c-T} \, \frac{1}{\,\sqrt{\,\xi^2+u(T,\, \xi)^2\,}+\xi\,} \, d\xi
=N_0 \int_{\varepsilon}^{\hslash\omega_D}
 \frac{\, v(\xi) \,}{\xi} \, d\xi.
\]
We can deal with the second and third terms similarly. We get
\begin{eqnarray*}
& & -N_0 \lim_{T \uparrow T_c} \int_{\varepsilon}^{\hslash\omega_D}
  \frac{\, u(T,\, \xi)^2 \,}{T_c-T} \, \frac{1}{\,\sqrt{\,\xi^2+u(T,\, \xi)^2\,}\,} \,
 \tanh \frac{\,\sqrt{\,\xi^2+u(T,\, \xi)^2\,}\,}{2T} \, d\xi \nonumber \\
&=& -N_0 \int_{\varepsilon}^{\hslash\omega_D}
  \frac{\, v(\xi) \,}{\xi} \tanh \frac{\, \xi \,}{2T_c} \, d\xi \nonumber \\
\end{eqnarray*}
and
\[
4N_0 \lim_{T \uparrow T_c} T \int_{\varepsilon}^{\hslash\omega_D} \frac{1}{\, T_c-T \,}
 \ln \frac{\, 1+e^{ -\sqrt{\,\xi^2+u(T,\, \xi)^2\,}/T } \,}{  1+e^{-\xi/T}  } \, d\xi
= -2N_0 \int_{\varepsilon}^{\hslash\omega_D}  \frac{\, v(\xi) \,}{\xi} \,
 \frac{1}{\, e^{\xi/T_c}+1 \,} \, d\xi \,.
\]

We thus see that $\Psi$ is differentiable at $T=T_c$ and that
\[
\lim_{T \uparrow T_c} \frac{\, \Psi(T_c)-\Psi(T) \,}{T_c-T}=0.
\]
\end{proof}

A straightforward calculation gives the following.
\begin{lemma}\label{lm:function-g}
Let $g$ be as in \eqref{eqn:function-g}. Then $g \in C^1[0,\,\infty)$, and
\[
g(\eta)<0 \quad (\eta \geq 0), \quad g'(0)=0, \quad \lim_{\eta \to \infty} g(\eta)= \lim_{\eta \to \infty} g'(\eta)=0.
\]
\end{lemma}

\begin{lemma}\label{lm:gapform}
Let $\Psi$ be as in \eqref{eqn:thermopotential}. Then $\Psi \in C^2[\tau,\, T_c]$, and
\[
\frac{\,\partial^2 \Psi \,}{\partial T^2}(T_c)=\frac{N_0}{\, 8\, T_c^2\,} \, \int_{\varepsilon/(2T_c)}^{\hslash\omega_D/(2T_c)}
 v(2\, T_c\, \eta)^2 g(\eta) \, d\eta \quad  (<0).
\]
\end{lemma}

\begin{proof}
A straightforward calculation gives that $(\partial \Psi/\partial T)$ is differentiable on $[\tau,\, T_c)$ and that $(\partial^2 \Psi/\partial T^2)$ is continuous on $[\tau,\, T_c)$. So it suffices to show that $(\partial \Psi/\partial T)$ is differentiable at $T=T_c$ and that $(\partial^2 \Psi/\partial T^2)$ is continuous at $T=T_c$ .
Note that $(\partial \Psi/\partial T)(T_c)=0$ by Lemma \ref{lm:psia}. Then
\begin{eqnarray}\label{eqn:psi4}
& & \quad \frac{\, \frac{\, \partial \Psi \,}{\,\partial T\,}(T_c)-\frac{\, \partial \Psi\,}{\,\partial T\,}(T) \,}{T_c-T}   \\ \nonumber
&=& -N_0 \int_{\varepsilon}^{\hslash\omega_D}
 \frac{\, u(T,\, \xi)^2 \, u(T,\, \xi) \frac{\, \partial u \,}{\, \partial T\,}(T,\,\xi) \,}{\,(T_c-T)(\xi^2+u(T,\,   \xi)^2)\,}
 \left\{
 \frac{1}{\, 2T \cosh^2 \frac{\,\sqrt{\,\xi^2+u(T,\, \xi)^2\,}}{2T} \,}
 -\frac{\,  \tanh \frac{  \,\sqrt{\,\xi^2+u(T,\, \xi)^2\,}  \,}{  2T  }   \,}{\,  \sqrt{\,\xi^2+u(T,\, \xi)^2\,}  \,}
 \right\} 
 d\xi \\ \nonumber
& & +N_0 \int_{\varepsilon}^{\hslash\omega_D}
 \frac{\, u(T,\, \xi)^2 \,}{T_c-T} \, 
 \frac{1}{\, 2T^2 \cosh^2 \frac{\,\sqrt{\,\xi^2+u(T,\, \xi)^2\,}}{2T} \,} \, d\xi \nonumber \\
& & +4N_0 \int_{\varepsilon}^{\hslash\omega_D} \frac{1}{\, T_c-T \,} \,
 \ln \frac{\, 1+e^{ -\sqrt{\,\xi^2+u(T,\, \xi)^2\,}/T } \,}{  1+e^{-\xi/T}  } \, d\xi \nonumber \\
& & +4N_0 \int_{\varepsilon}^{\hslash\omega_D} \frac{1}{\, T_c-T \,} \,
 \left\{
 \frac{\sqrt{\,\xi^2+u(T,\, \xi)^2\,}/T}{\, e^{\sqrt{\,\xi^2+u(T,\, \xi)^2\,}/T }+1 \,}
  -\frac{\xi/T}{\, e^{\xi/T}+1 \,} 
 \right\} \, d\xi. \nonumber \\ \nonumber
\end{eqnarray}
By (C2) of Condition (C), for an arbitrary $(0<) \varepsilon_1<1$, there is a $\delta>0$ such that $| T_c-T |<\delta$ implies
\begin{eqnarray*}
\left| \frac{\, u(T,\, \xi)^2 \,}{T_c-T} \, 
 \frac{\, u(T,\, \xi) \frac{\, \partial u \,}{\, \partial T\,}(T,\,\xi) \,}{\,\xi^2+u(T,\, \xi)^2\,} \right|
&<& \frac{\, \left( v(\xi)+T_c\,\varepsilon_1\right)^2 \,}{2\xi^2} < \frac{\, \left( v(\xi)+T_c \right)^2 \,}{2\xi^2}. \\
\end{eqnarray*}
The Lebesgue dominated convergence theorem therefore implies that the first term on the right side of \eqref{eqn:psi4} becomes
\[
\frac{\, N_0 \,}{2} \int_{\varepsilon}^{\hslash\omega_D}
 \frac{\, v(\xi)^2 \,}{\xi^2} \,  \left\{
 \frac{1}{\, 2T_c \cosh^2 \frac{\, \xi \,}{2T_c} \,}-\frac{\,  \tanh \frac{  \, \xi  \,}{2T_c}   \,}{\xi} 
   \right\} \, d\xi
\]
as $T \uparrow T_c$ . Similarly, the rest on the right side of \eqref{eqn:psi4} becomes $0$ as $T \uparrow T_c$ . We thus find that $(\partial \Psi/\partial T)$ is differentiable at $T=T_c$ and that
\[
\frac{\,\partial^2 \Psi \,}{\partial T^2}(T_c)=\frac{\, N_0 \,}{2} \int_{\varepsilon}^{\hslash\omega_D}
 \frac{\, v(\xi)^2 \,}{\xi^2} \,  \left\{
 \frac{1}{\, 2T_c \cosh^2 \frac{\, \xi \,}{2T_c} \,}-\frac{\,  \tanh \frac{  \, \xi  \,}{2T_c}   \,}{\xi} 
   \right\} \, d\xi.
\]
Continuity of $(\partial^2 \Psi/\partial T^2)$ at $T=T_c$ follows immediately.
\end{proof}

Our proof of Theorem \ref{thm:2ndorderphase} is complete. Moreover, Proposition \ref{prp:gapheat} follows immediately from Remark \ref{rmk:gapcv} and Lemma \ref{lm:gapform}.

\end{document}